\documentclass[twocolumn,floatfix,prb,aps,showpacs,amsmath]{revtex4-1}
\usepackage{graphicx,amsmath,amssymb,color,xcolor}
\usepackage{nicefrac}
\usepackage[titletoc,title]{appendix}

\newcommand{\be}{\begin{equation}}
\newcommand{\ee}{\end{equation}}

\newcommand{\ba}{\begin{eqnarray}}
\newcommand{\ea}{\end{eqnarray}}

\begin{document}

\title{A non-Abelian fractional quantum Hall state at $3/7$ filled Landau level}
\author{W. N. Faugno$^{1,2}$, J. K. Jain$^1$, and Ajit C. Balram$^{3}$}
\affiliation{$^1$Department of Physics, 104 Davey Lab, Pennsylvania State University, University Park, Pennsylvania 16802, USA}
\affiliation{$^2$Institut de Physique Theorique, Universit\'e Paris-Saclay, CNRS, CEA, 91190 Gif sur Yvette, France}
\affiliation{$^{3}$Institute of Mathematical Sciences, HBNI, CIT Campus, Chennai 600113, India}
\date{\today}

\begin{abstract} 
We consider a non-Abelian candidate state at filling factor $\nu=3/7$ state belonging to the parton family. We find that, in the second Landau level of GaAs (i.e. at filling factor $\nu=2+3/7$), this state is energetically superior to the standard Jain composite-fermion state and also provides a very good representation of the ground state found in exact diagonalization studies of finite systems. This leads us to predict that \emph{if} a fractional quantum Hall effect is observed at $\nu=3/7$ in the second Landau level, it is likely to be described by this new non-Abelian state. We enumerate experimentally measurable properties that can verify the topological structure of this state. 
\end{abstract}
\maketitle

\section{Introduction}

It is a remarkable, and perhaps somewhat surprising, fact that the fractional quantum Hall effect (FQHE) in the second Landau level (LL) of semiconductor quantum wells often has a different origin than that in the lowest LL (LLL). A compressible state is seen at the half-filled LLL ($\nu=1/2$), whereas FQHE is seen at $\nu=5/2$, i.e. the half-filled second LL (SLL )~\cite{Willett87}. The former is well understood as a Fermi sea of composite fermions (CFs)~\cite{Halperin93, Jain07, Halperin20}, whereas the latter is believed to be a paired state of composite fermions, described by the Moore-Read Pfaffian wave function~\cite{Moore91, Read00} or its particle-hole conjugate~\cite{Lee07, Levin07}. The FQHE at $\nu=2/5$ in the LLL is a Jain CF state~\cite{Jain89}, whereas at $\nu=2/5$ in the second LL (i.e. $\nu=12/5$) it is not~\cite{Wojs09,Bonderson12,Sreejith13}; it is likely described by the particle-hole conjugate of a Read-Rezayi wave function~\cite{Read99,Rezayi09,Zhu15,Mong15,Pakrouski16}. Even the applicability of the Laughlin wave function~\cite{Laughlin83} to the $1/3$ FQHE in the second LL ($\nu=7/3$) has been debated~\cite{Ambrumenil88,Balram13b,Johri14,Peterson15,Kleinbaum15,Jeong16,Balram19a,Yoo19}. 

In recent years, Balram and collaborators have demonstrated that some of Jain's parton states~\cite{Jain89b} are reasonable candidates for the second LL FQH states, such as those at $\nu=5/2$~\cite{Balram18}, $\nu=7/3$~\cite{Balram19a}, $\nu=12/5$~\cite{Balram19}, and $\nu=2+6/13$~\cite{Balram18a,Balram20}. (We stress here that the $n=1$ LL of monolayer graphene behaves differently than the SLL of ordinary semiconductors such as GaAs. The physics of the $n=1$ LL of monolayer graphene is similar to that of the LLL of GaAs, described very well in terms of the CF theory~\cite{Balram15c}.) Certain other states from the parton construction have been proposed for higher graphene LLs~\cite{Wu17, Kim19} and wide quantum wells~\cite{Faugno19}. Many of these states support non-Abelian quasiparticles, as shown by Wen~\cite{Wen91, Wen92b}. 

In this article, we study the competition between two members of the parton family, labeled by $\bar{3}\bar{3}111$ and $311$ (meaning explained below), both of which are candidate states at $\nu=3/7$. The $311$ state represents the familiar integer quantum Hall (IQH) state of composite fermions~\cite{Jain89}, has Abelian quasiparticles, and is known to describe the $3/7$ FQHE in the LLL. In contrast, the $\bar{3}\bar{3}111$ state is believed to support non-Abelian quasiparticles that have sufficiently rich braid statistics to allow, in principle, universal topological quantum computation (similar to what is believed for the $12/5$ FQHE). Our principal result is to show that in the SLL of GaAs, i.e. at $\nu=2+3/7$, the $\bar{3}\bar{3}111$ state is variationally superior to the $311$ state, and is also an excellent representation of the ground state found in exact diagonalization studies for finite systems. 

It is worth remarking on the current experimental status of the nature of the state at $\nu=2+3/7$. An earlier experimental work~\cite{Choi08} reported a shoulder in $R_{xx}$ near $\nu=2+3/7$ in GaAs quantum wells, suggesting an incipient FQHE, but experiments in better quality samples and at lower temperatures~\cite{Kumar10, Deng12, Shingla18} show an $R_H=h/(2e^2)$ IQH plateau (``re-entrant IQH state") at $\nu=2+3/7$, indicating the formation of presumably a bubble crystal that is pinned by the disorder. Evidence for $10/7$ and $11/7$ FQHE has recently been seen in bilayer graphene (BLG)~\cite{Zhu20}, although it is unclear at the moment if the LL orbital hosting this state is more similar to the lowest or the second LL of semiconductor quantum wells. (The band structure and hence the Landau level structure of bilayer graphene is known under some assumptions. The ``zeroth" LL consists of doublets labeled as $|0\rangle$ and  $|1\rangle$. The state $|0\rangle$ is identical to the LLL state of GaAs. The state $|1\rangle$ is an ``admixture" of the LLL and SLL states of GaAs, where the relative importance of the two components is controlled by parameters such as the magnetic field. As the magnetic field is increased, the state $|1\rangle$ goes from being similar to the SLL of GaAs to being similar to the $n=1$ LL of monolayer graphene; the physics of the latter is well described in terms of weakly interacting composite fermions~\cite{Balram15c}, producing the standard Jain CF state at $\nu=3/7$.) A bump in the incompressibility has also been observed in BLG in the $n=1$ LL at $4/7$ filling~\cite{Zibrov17}. If this indeed is confirmed as an FQHE state, the hole partner of our $3/7$ parton state should be the leading candidate for its explanation.

Our numerical studies do not rule out the possibility that the actual ground state at $\nu=2+3/7$ is an FQHE liquid. This possibility would be consistent with experiments if the FQHE state is masked by disorder, given that disorder favors a crystal over a liquid; in that scenario, an FQHE at $\nu=2+3/7$ would reveal itself in still better quality samples. The other possibility, of course, is that the true ground state at $\nu=2+3/7$ is a bubble crystal. However, our numerical results suggest that the incompressible $\bar{3}\bar{3}111$ liquid state is very competitive and might be stabilized by changing the effective interaction between electrons, which can be accomplished by changing the width of the quantum well and/or the density, or LL mixing, or by screening the interaction with the help of a nearby metallic layer. If an FQHE is observed at $\nu=2+3/7$, the considerations of this article should be relevant.

\section{The Parton Construction}

The parton construction~\cite{Jain89b} provides a theoretical framework for constructing new candidate FQH states from known QH states. In the parton construction, one considers breaking the electrons into $m$ species of fictitious particles called partons, with the parton species labeled by $\lambda$. Because the density of each parton species must be the same as the electron density [given by $\rho=e\nu B/(hc)$, where $B$ is the external magnetic field], we must have $e\nu=e_\lambda \nu_\lambda$, indicating that the charge of the $\lambda$-parton is given by $e_\lambda = \nu/\nu_\lambda$ in units of the electron charge $(-e)$. Because the parton charges must add to the electron charge, we have $\sum_{\lambda=1}^m e_\lambda = -e$. It follows that electron filling $\nu$ is related to the parton fillings $\{\nu_\lambda\}$ as $\nu = [\sum_{\lambda=1}^m \nu^{-1}_\lambda]^{-1}$. A candidate incompressible state is produced when each parton species occupies an incompressible state, in particular an IQH state with filling $\nu_\lambda = n_\lambda$. We label the incompressible parton states by their integer fillings $n_1n_2...n_m$. This leads to trial wave functions of the form
\begin{equation}
\Psi^{n_1n_2...n_m} = \mathcal{P}_{\rm LLL}\prod_{\lambda=1}^{m} \Phi_{n_{\lambda}}\left(\{z_{k}\}\right),
\end{equation}
where $\Phi_n$ is the Slater determinant wave function for $n$ filled Landau levels of noninteracting particles, $z_k = x_k - i y_k$ are the complex coordinates of the $k$th electron, and $\mathcal{P}_{\rm LLL}$ is the lowest Landau level projector. Note that all the partons have the same coordinate as their parent electrons. This ``gluing'' procedure removes the artificial degrees of freedom introduced by breaking the electrons into partons. We also introduce the notation $\bar{n}$ to denote negative fillings (i.e., negative magnetic field), which correspond to factors of $\Phi_{\bar{n}}=\Phi_{n}^*$ in the trial wave functions. Parton states can be Abelian or non-Abelian. The Abelian Jain CF wave functions~\cite{Jain89} are a subset of the parton theory; they are states of the form $n11\cdots$. States with repeated factors of $\Phi_{n_{\lambda}}$, where $|n_{\lambda}|\geq 2$, are non-Abelian~\cite{Blok90, Blok90b, Wen91}.

\subsection*{A non-Abelian state at $\nu=3/7$}
At filling factor $\nu = 3/7$, the parton theory provides a non-Abelian candidate state $\bar{3}\bar{3}111$ described by the wave function
\begin{equation}
\Psi^{\bar{3}\bar{3}111}_{\nu=3/7}=\mathcal{P}_{\rm LLL}  [\Phi_{\bar{3}}]^{2}\Phi_1^3 = \frac{[\Psi^{\rm CF}_{3/5}]^{2}}{\Phi_{1}}.
\label{eq: parton_bar3bar3111} 
\end{equation}
In the second equality above, we have used $\Psi^{\rm CF}_{3/5}=\mathcal{P}_{\rm LLL}  
\Phi_{\bar{3}} \Phi_1^2$ to define the LLL projection of $\Psi^{\bar{3}\bar{3}111}_{\nu=3/7}$ in a specific fashion; past work has indicated that different ways of accomplishing LLL projection yield very similar wave functions, and in particular do not alter the topological nature of the state~\cite{Balram16b}. A nice feature of the $\bar{3}\bar{3}111$ wave function as expressed in Eq.~(\ref{eq: parton_bar3bar3111}) is that it can be evaluated for large system sizes which allow a reliable extrapolation of its thermodynamic energy. This relies on the fact that the  $n11$ states can be evaluated for hundreds of electrons using the Jain-Kamilla (JK) method of projection~\cite{Jain97b, Moller05, Jain07, Davenport12, Balram15a}. For the $\bar{n}11$ states, that involve reverse flux attachment (i.e., negative filling factors), it is time-consuming to evaluate the states for systems sizes larger than $50$ electrons since the wave function calculation requires high-precision arithmetic which slows down the computation considerably; nonetheless, as we see below, reliable thermodynamic limits can be obtained with systems accessible to numerical evaluation.  We note that the wave functions we consider are always constructed in the LLL since they can readily be evaluated in this form. The second LL physics will be simulated in the LLL by using an effective interaction whose Haldane pseudopotentials in the LLL are very nearly the same as the second LL pseudopotentials of the Coulomb interaction.

The $\bar{3}\bar{3}111$ state has a Wen-Zee~\cite{Wen92} shift $\mathcal{S}=-3$ in the spherical geometry, defined below. Due to repeated factors of $\bar{3}$, the $\bar{3}\bar{3}111$ state is a non-Abelian state hosting quasiparticles whose non-Abelian fusion rules are described by an $SU(2)_{-3}$ Chern-Simons (CS) theory~\cite{Wen91, DiFrancesco97, Balram19}. The braiding properties of these so-called Fibonacci anyons are rich enough to potentially carry out \emph{universal} fault-tolerant topological quantum computation~\cite{Nayak08, Rowell09}. 

\section{Numerical Results}

We perform variational Monte Carlo (VMC) and exact-diagonalization (ED) calculations in the spherical geometry~\cite{Haldane83} wherein $N$ electrons are confined to the surface of a sphere. The radial magnetic field is generated by a magnetic monopole placed at the center of the sphere. The strength of the magnetic monopole is denoted by the integer $2Q$, which produces a magnetic flux of $2Q\phi_0$, where $\phi_0=hc/e$ is the magnetic flux quantum. The radius of the sphere is given by $\sqrt{Q}\ell$ where $\ell = \sqrt{\hbar c/(eB)}$ is the magnetic length. On the sphere, the single-particle orbitals in the Landau level indexed by $n=0,1,\cdots$ are eigenstates of angular momentum operator with eigenvalue $l=|Q|+n$, and can be labeled by the $z$-component of their orbital momentum, $l_z$, which ranges from $-(|Q|+n)$ to $|Q|+n$. For incompressible states, the strength of the magnetic monopole is the sum of the effective magnetic monopoles required to create each parton species's IQH effect (IQHE), i.e. $2Q = \sum_\lambda 2Q_\lambda^* = \sum_\lambda \left(N/n_\lambda - n_\lambda\right)$. In constructing states on the sphere, we note the shift~\cite{Wen92} is given by $\mathcal{S} = \nu^{-1}N - 2Q$. The incompressible states are uniform on the sphere i.e., have total orbital angular momentum $L=0$. 

Throughout this work, we assume that the magnetic field is sufficiently large to fully spin polarize the electrons. We also neglect the effects of LL mixing and disorder. We will assume zero width for most of our work, but will also consider how finite quantum well width affects our results. In the LLL, we use the bare Coulomb interaction to calculate the energy of the zero-width system. For the $n=1$ LL of GaAs and graphene, we use the effective interactions employed in Refs.~\cite{Toke08} and \cite{Balram15c}, respectively, to simulate the physics of the $n=1$ LL in the LLL. 

\subsection{Ground state}

\begin{figure*}[htpb]
\begin{center}
\includegraphics[width=0.31\textwidth,height=0.23\textwidth]{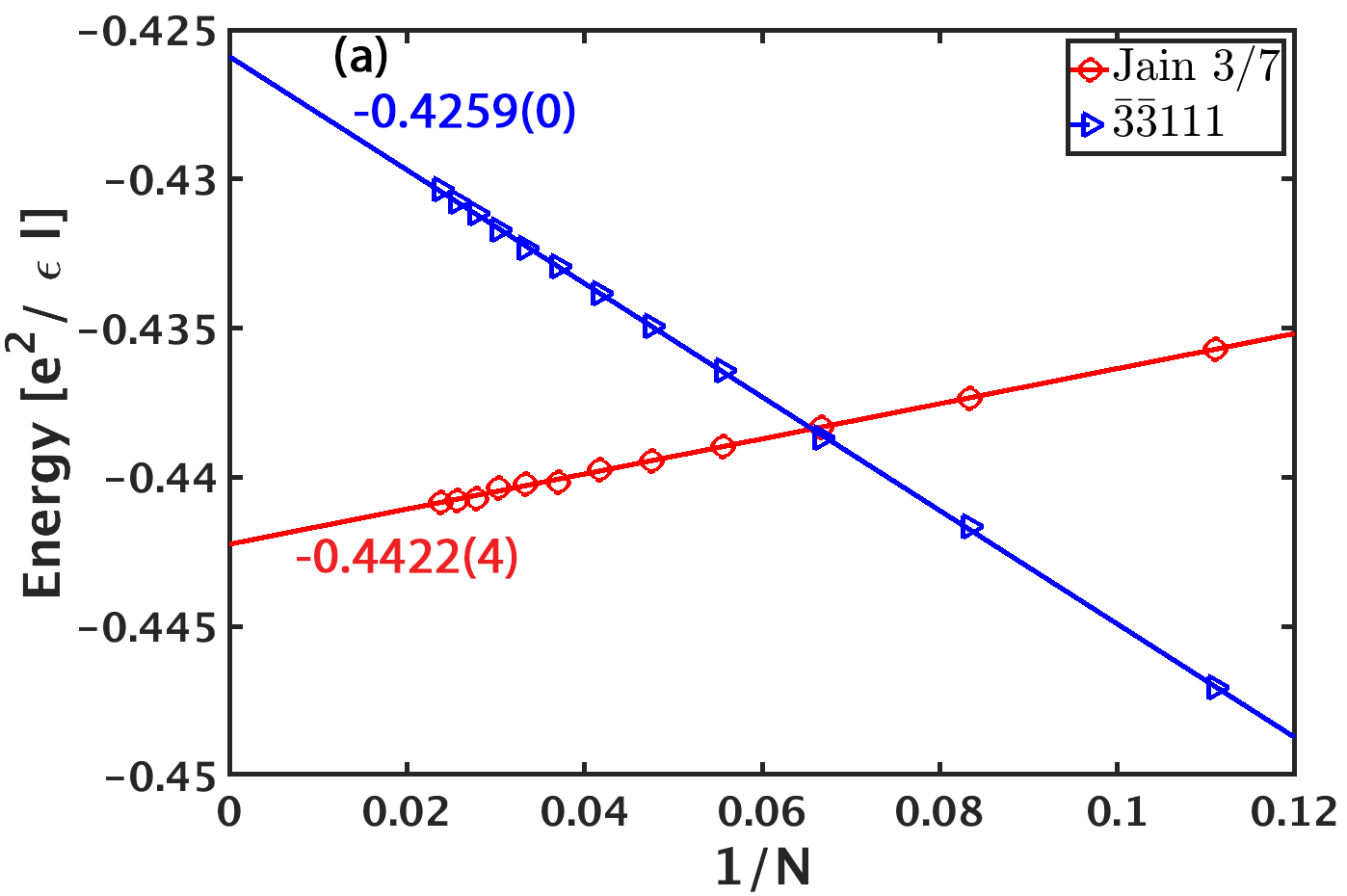} 
\includegraphics[width=0.31\textwidth,height=0.23\textwidth]{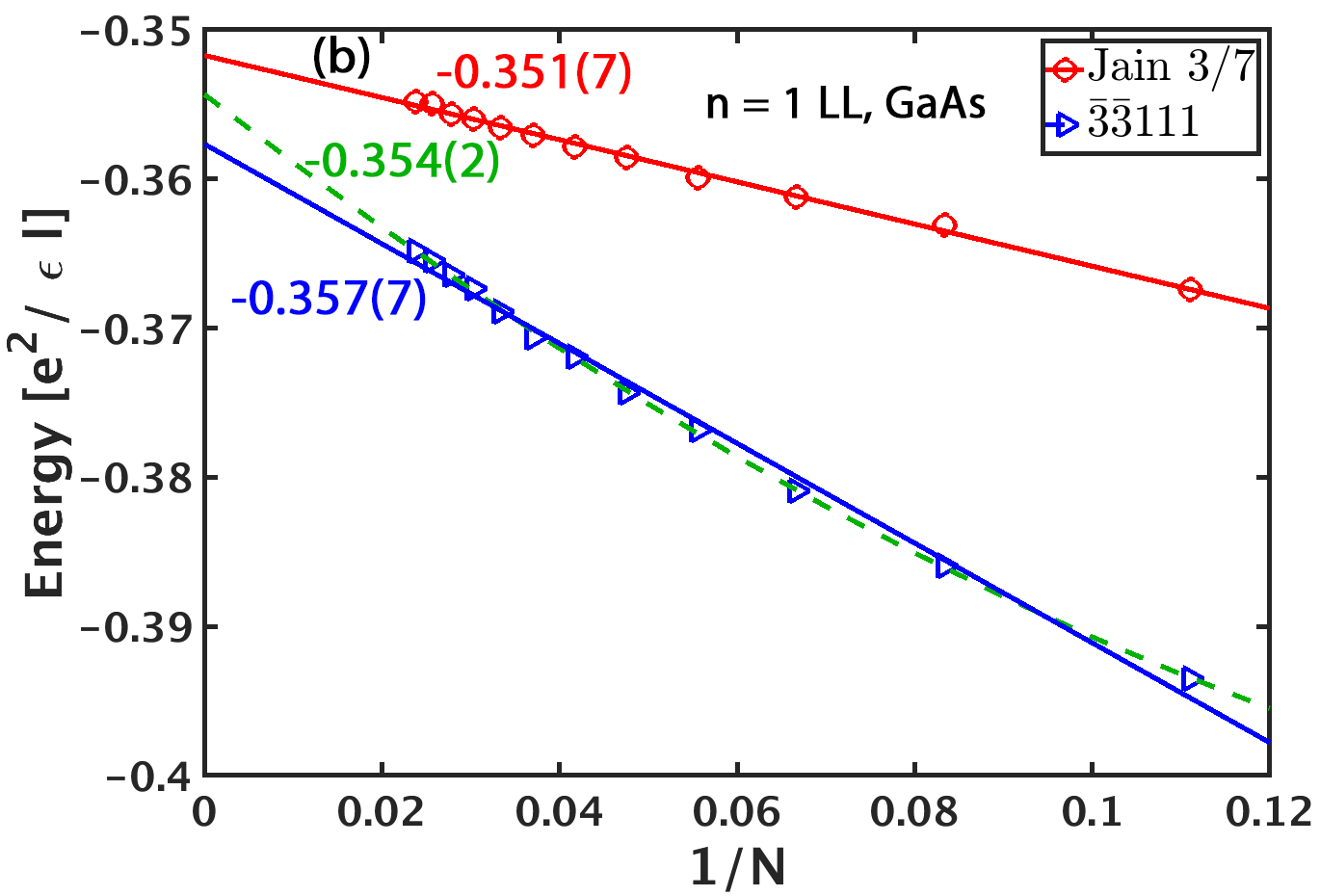} 
\includegraphics[width=0.31\textwidth,height=0.23\textwidth]{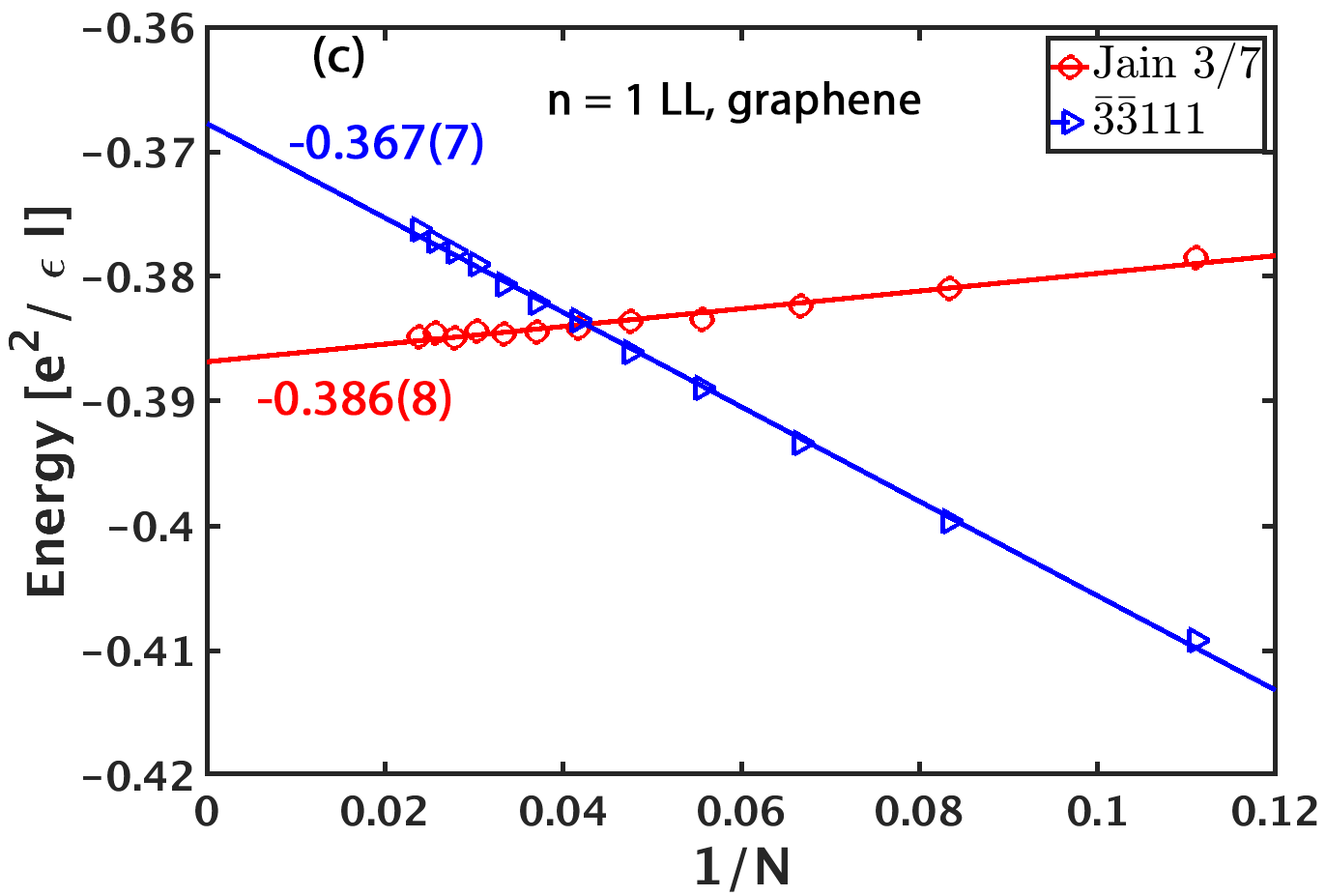} 
\caption{(color online) Thermodynamic extrapolations of the per-particle Coulomb energies for the $311$ and (red circles) and the $\bar{3}\bar{3}111$ states (blue triangles) at $\nu=3/7$. Panels (a), (b) and (c) show energies in the $n=0$ LL, $n=1$ LL of GaAs, and in the $n=1$ LL of monolayer graphene, respectively. The extrapolated energies, obtained from a linear fit in $1/N$, are quoted in Coulomb units of $e^2/(\epsilon\ell)$ in the plot, and the number in the parentheses indicates the uncertainty in the linear fit. These energies include contributions of the electron-background and background-background interaction, and are density-corrected~\cite{Morf86}. In panel (b), we have added a quadratic fit in $1/N$ (green dashed line) to demonstrate that the ordering of states does not depend on the extrapolation method. The Coulomb energies for the Jain 311 state in the $n=0$ and $n=1$ LLs of monolayer graphene have been reproduced from Refs.~\cite{Balram17} and~\cite{Balram15c}.}
\label{fig: extrapolations_energies_3_7}
\end{center}
\end{figure*}

In Fig.~\ref{fig: extrapolations_energies_3_7} we compare the VMC energies of the $\bar{3}\bar{3}111$  and the $311$ wave functions at $\nu=3/7$ in the LLL and the $n=1$ LLs of GaAs and monolayer graphene. These energies include the electron-electron, electron-background, and background-background contributions [the last two collectively contribute $-N^2/(2\sqrt{Q}\ell)$ to the total energy]; we further multiply these energies by $\sqrt{2Q\nu/N}$, which corrects for the finite size deviation of the density from its thermodynamic value and thus minimizes finite-size effects~\cite{Morf86}. All energies are given in units of $e^2/(\epsilon \ell)$ where  $\epsilon$ is the dielectric constant of the host material. As anticipated, in the LLL the $311$ state has lower energy than the $\bar{3}\bar{3}111$ state. In contrast, in the SLL of GaAs, we find that the $\bar{3}\bar{3}111$  state has lower energy. In the $n=1$ LL of monolayer graphene, we find the $311$ state has lower energy, consistent with the fact that FQHE states in the $n=1$ LL of monolayer graphene conform to the CF paradigm~\cite{Amet15, Balram15c, Zeng18}. Under our working assumption of neglecting the effects of finite width and LL mixing, the results for the $n=0$ LL of graphene are identical to those in the LLL of GaAs.

%%%%%%%%%%%%%%%%%%%%%%%%%%%%%%%%%%%%%%%%%%%%%%%%%%%%%%%%%%%%%%%%%%%%%%%%%%%%%%
%%%%%%%%%%%%%%%%%%%%%%%%%%%%%%%%%%%%%%%%%%%%%%%%%%%%%%%%%%%%%%%%%%%%%%%%%%%%%%

Next, we present results obtained from ED at the appropriate flux of $2Q =7N/3+3$. The exact ground states in the SLL for the three smallest systems of $N=9,12,15$ electrons are all uniform, i.e., have $L=0$. The next system size of $N=18$ electrons, which has a Hilbert space dimension of above $25$ billion, is not accessible to ED. The exact SLL Coulomb ground state has an overlap of $0.92$ with the $\bar{3}\bar{3}111$ state for $N=9$. (For this purpose, we have obtained an exact Fock-space representation of the $\bar{3}\bar{3}111$ state using the method described in Ref.~\cite{Sreejith11}.) This overlap is small compared to those we encounter for $n11$ states in the LLL, but quite high for typical comparisons in the SLL; for example, the $7/3$ Coulomb ground state for $N=9$ electrons has an overlap of $0.48$ with the Laughlin $111$ state. The calculation of the Fock-space representation of the $\bar{3}\bar{3}111$ state with $N\geq12$ is beyond our computational reach, which precludes a determination of its overlap with the exact ground state for these systems. 

\begin{figure}[htpb]
\begin{center}
\includegraphics[width=0.47\textwidth,height=0.26\textwidth]{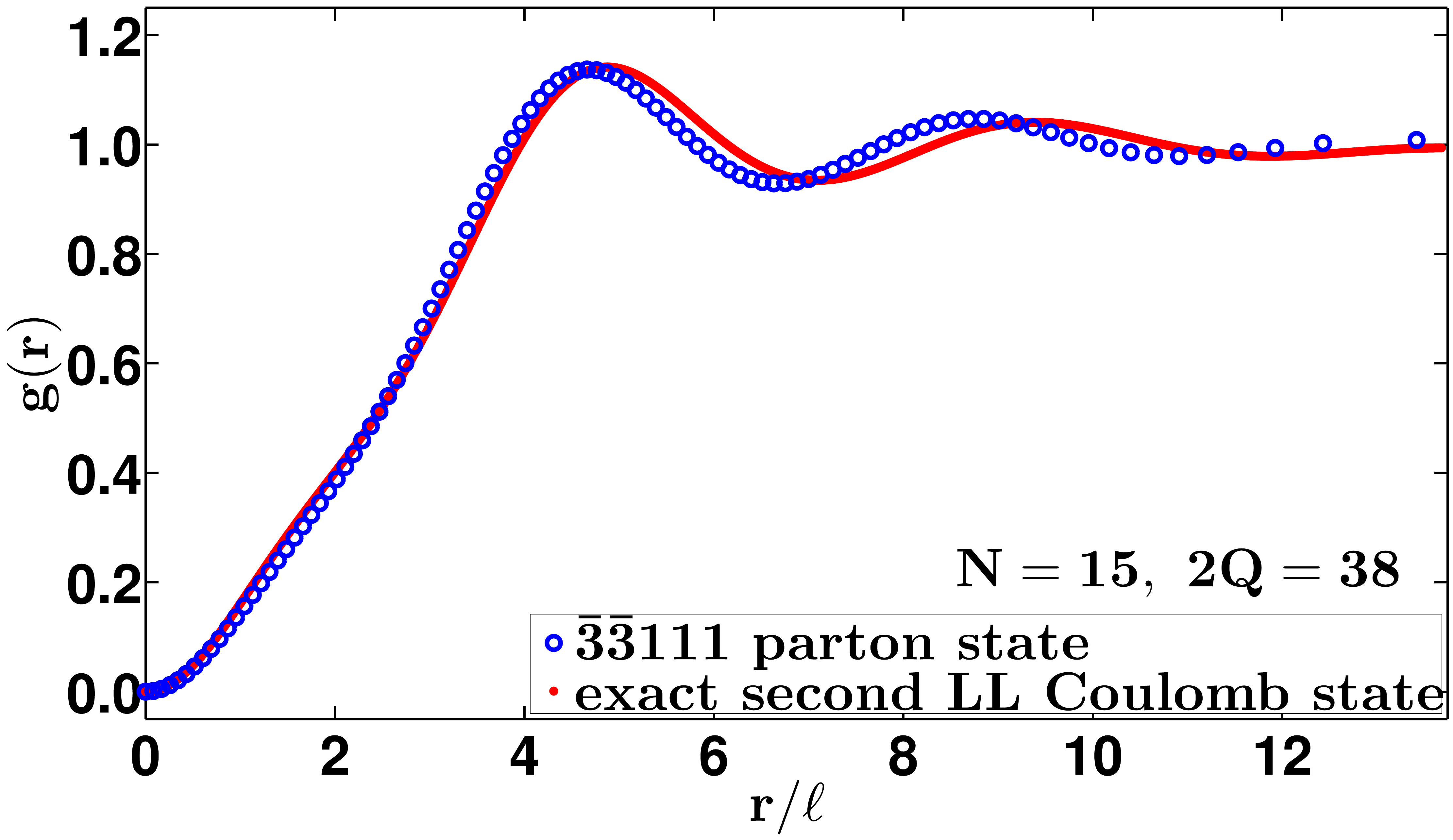} 
\caption{(color online) The pair correlation function $g(r)$ as a function of the arc distance $r$ on the sphere for the exact second Landau level Coulomb ground state (red filled dots), and the $\bar{3}\bar{3}111$ state of Eq.~(\ref{eq: parton_bar3bar3111}) (blue open circles) for $N=15$ electrons at a flux of $2Q=38$.}
\label{fig: pair_correlations_3_7}
\end{center}
\end{figure}

We next compare the pair-correlation function of the exact SLL Coulomb ground state with that of the $\bar{3}\bar{3}111$ state for $N=15$ electrons (see Fig.~\ref{fig: pair_correlations_3_7}). The pair-correlation functions of both states show oscillations that decay at long distances, which is a typical characteristic of an incompressible state~\cite{Kamilla97, Balram15b}. Moreover, both states show a ``shoulder''-like feature in the pair-correlation function at short distances, which is a signature of clustering in non-Abelian states~\cite{Read99, Hutasoit16}. Furthermore, these two pair-correlation functions are in remarkably good agreement with each other. To test whether the results depend sensitively on the choice of the pseudopotentials, we have also evaluated the exact ground state in the SLL with the truncated planar disk pseudopotentials for $N=15$ electrons. The overlap between the ground states obtained using the disk and spherical pseudopotentials is $0.9792$ which shows that the two ground states are quite close to each other.

%%%%%%%%%%%%%%%%%%%%%%%%%%%%%%%%%%%%%%%%%%%%%%%%%%%%%%%%%%%%%%%%%%%%%%%%%%%%%%
%%%%%%%%%%%%%%%%%%%%%%%%%%%%%%%%%%%%%%%%%%%%%%%%%%%%%%%%%%%%%%%%%%%%%%%%%%%%%%

For the $N=15$ system, the exact energy for the effective interaction we use to simulate the physics of the $n=1$ LL in the LLL is -0.3826. In comparison, the $\bar{3}\bar{3}111$ state has an energy of -0.3809(1) for the same interaction (The number in the parenthesis is the statistical uncertainty in the Monte Carlo estimate of the energy of the $\bar{3}\bar{3}111$ state.) This level of agreement is on-par with that of other trial states in the second Landau level~\cite{Balram20}. For completeness, we have also evaluated the exact LLL Coulomb ground state for the same system. The ground state in the LLL is not uniform (has $L=4$), which indicates that the natures of the ground states in the two LLs are quite different from each other for this system.

These studies show that the $\bar{3}\bar{3}111$ state is variationally better than the 311 state in the SLL, and it also provides a very good approximation for the exact ground state for systems accessible to ED. These facts establish the plausibility of the $\bar{3}\bar{3}111$ state for FQHE at $\nu=2+3/7$ in the SLL or GaAs.

\subsection{Excitation gaps}

\begin{figure}[htpb]
\includegraphics[width=0.47\textwidth]{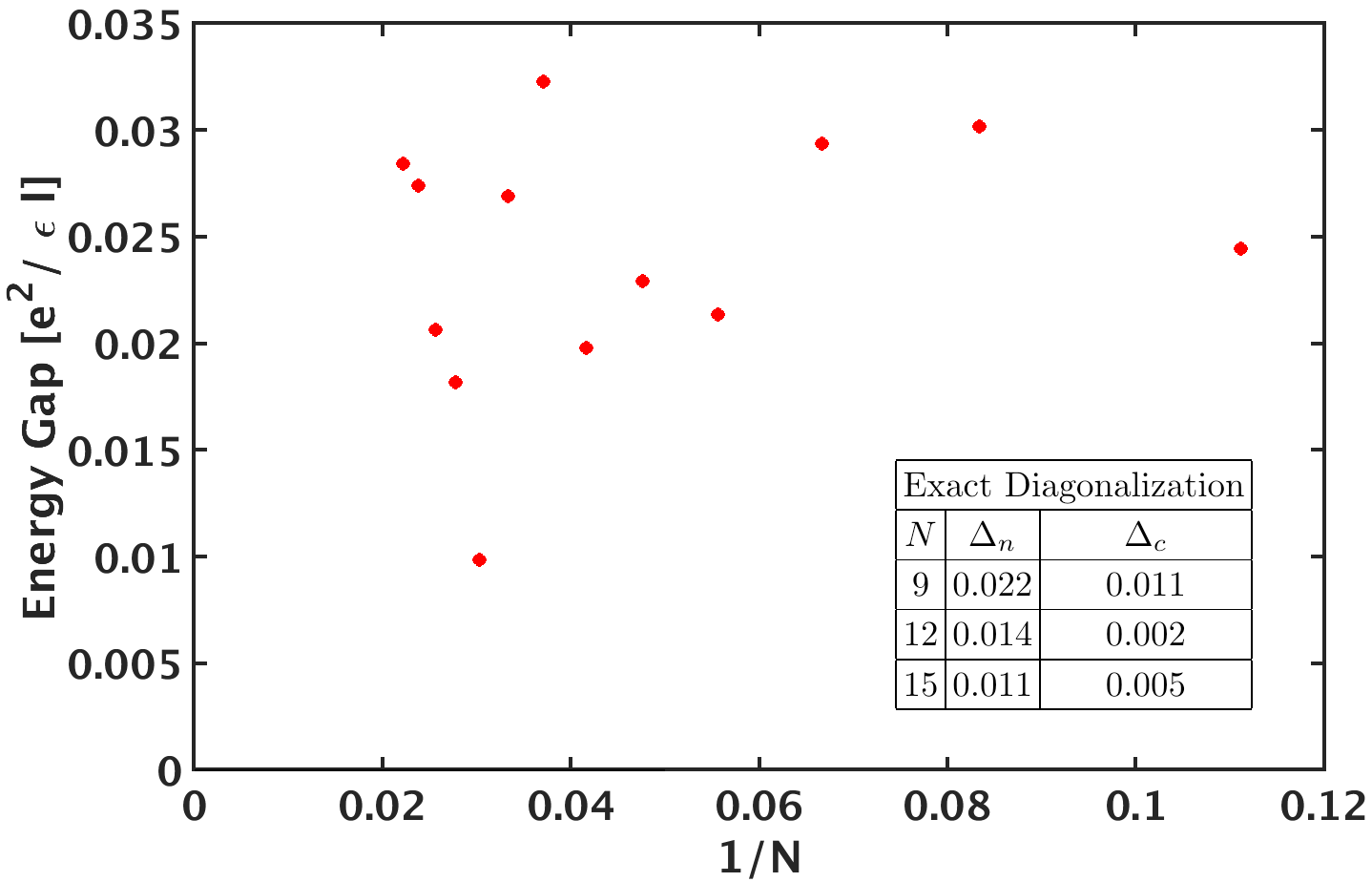}
\caption{(color online) Charge gap $\Delta_{c}$ calculated from the difference in energies of the trial wave functions presented in Eqs.~(\ref{eq: wf_exciton_3_5}) and (\ref{eq: parton_bar3bar3111}). The gap shows strong finite-size fluctuations, precluding a clear extrapolation to the thermodynamic limit, but is finite for all $N$ considered. The lack of a clear trend with $1/N$ is likely the result of the large overlap, and thus a significant interaction, of the quasiparticle and quasihole for small system sizes (see text and Fig.~\ref{fig: density_exciton_3bar3bar111}). We estimate from this calculation that the charge gap should be of the order 0.01$e^2/(\epsilon \ell)$. For comparison, the inset shows the charge and neutral gap $\Delta_{n}$ (see text for definition) for small systems evaluated from exact diagonalization.}
\label{fig: VMCgap}
\end{figure}

We can extract the charge and neutral gaps for the three smallest systems that are accessible to exact diagonalization. The charge gap here is defined as $\Delta_{c}=[\mathcal{E}(2Q=7N/3+4)+\mathcal{E}(2Q=7N/3+2)-2\mathcal{E}(2Q=7N/3+3)]/3$, where $\mathcal{E}(2Q)$ is the ground state energy at flux $2Q$ and the factor of $3$ in the denominator accounts for the fact that the addition or removal of a single flux quantum produces three fundamental quasiholes or quasiparticles in the $\bar{3}\bar{3}111$ state. The neutral gap $\Delta_{n}$ is defined as the difference between the two lowest energies at the flux of $2Q = 7N/3+3$. The density-corrected charge and neutral gaps, which include the background contribution, for the individual systems are reported in the table inset in Fig.~\ref{fig: VMCgap}. The charge and neutral gaps do not fit well to a linear function of $1/N$ and thus we do not have a reliable estimate of them in the thermodynamic limit. Also, the neutral gap is larger than the charge gap for these systems which indicates strong finite-size effects in the SLL. (We expect the charge gap to be larger than or equal to the neutral gap in the thermodynamic limit.)

We have also attempted to estimate the charge gap via a VMC. To do so we create a quasihole-quasiparticle pair by promoting one particle in one factor of $\Phi_{\bar{3}}$ to the fourth $\Lambda$ level ($\Lambda$L). ($\Lambda$Ls are the Landau-like levels occupied by partons at their respective effective magnetic field.) We assign the $z$-component of the orbital angular momentum of the single-particle state for the quasihole in the 3rd $\Lambda$L to be at the maximal value, namely $l_z = (|Q_{-3}^*| + 2)$, while the single-particle state of the quasiparticle in the 4th $\Lambda$L is placed in the state with minimal $z$-component, $l_z = -(|Q_{-3}^*|+3)$. Thus, the quasiparticle and quasihole are as far separated as possible, to minimize the contribution of their interaction to the total energy. The excited state is modeled using the parton wave function
 \begin{equation}
  \Psi^{\rm exciton}_{3/7} = \mathcal{P}_{\rm LLL}[\Phi^{\rm exciton}_{3}]^{*}[\Phi_{3}]^{*}\Phi^{3}_{1}= \frac{\Psi^{\rm CF-exciton}_{3/5}\Psi^{\rm CF}_{3/5}}{\Phi_{1}}.
  \label{eq: wf_exciton_3_5}
 \end{equation}
where the superscript ``exciton'' refers to the creation of a quasiparticle and quasihole in this factor. 
The gap does not follow a linear fit in $1/N$, indicating strong finite size corrections, and we only roughly estimate it to be of order 0.01 $e^{2}/(\epsilon\ell)$ as shown in Fig.~\ref{fig: VMCgap}. The reason for strong finite-size effects is that, as we show below, the quasihole and quasiparticle are very large, and for the system sizes that we can access, they overlap significantly. (As shown below, we need to go to systems with 90 particles before we see well-separated quasiparticle and quasihole.) We have not attempted to evaluate the charge gap using the standard procedure of inserting or removing a flux quantum since doing so creates 3 quasiholes or 3 quasiparticles in the $\bar{3}\bar{3}111$ state, which would overlap strongly and thus the interactions between them cannot be neglected. As a result, we cannot place the particles on the sphere so that they are well separated for any reasonably sized system accessible by the JK projection.
 
\subsection{Stability of $\bar{3}\bar{3}111$ state}

\begin{figure*}[htpb]
\begin{center}
\includegraphics[width=0.47\textwidth,height=0.19\textwidth]{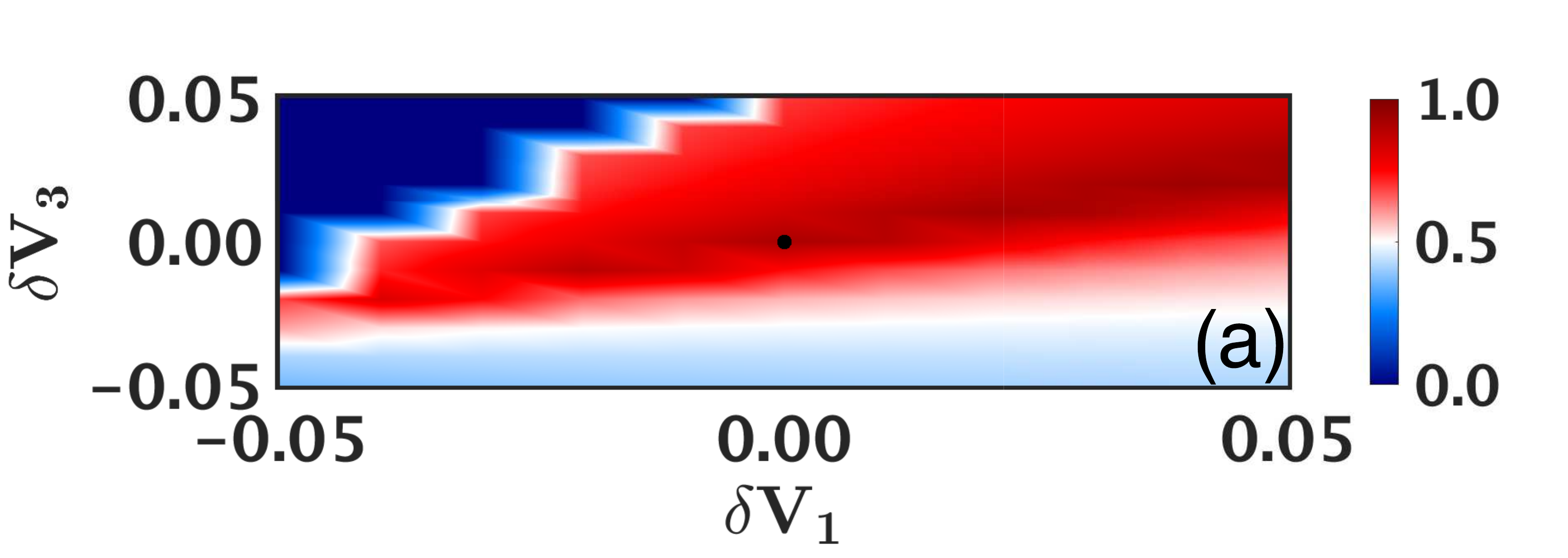}
\includegraphics[width=0.47\textwidth,height=0.19\textwidth]{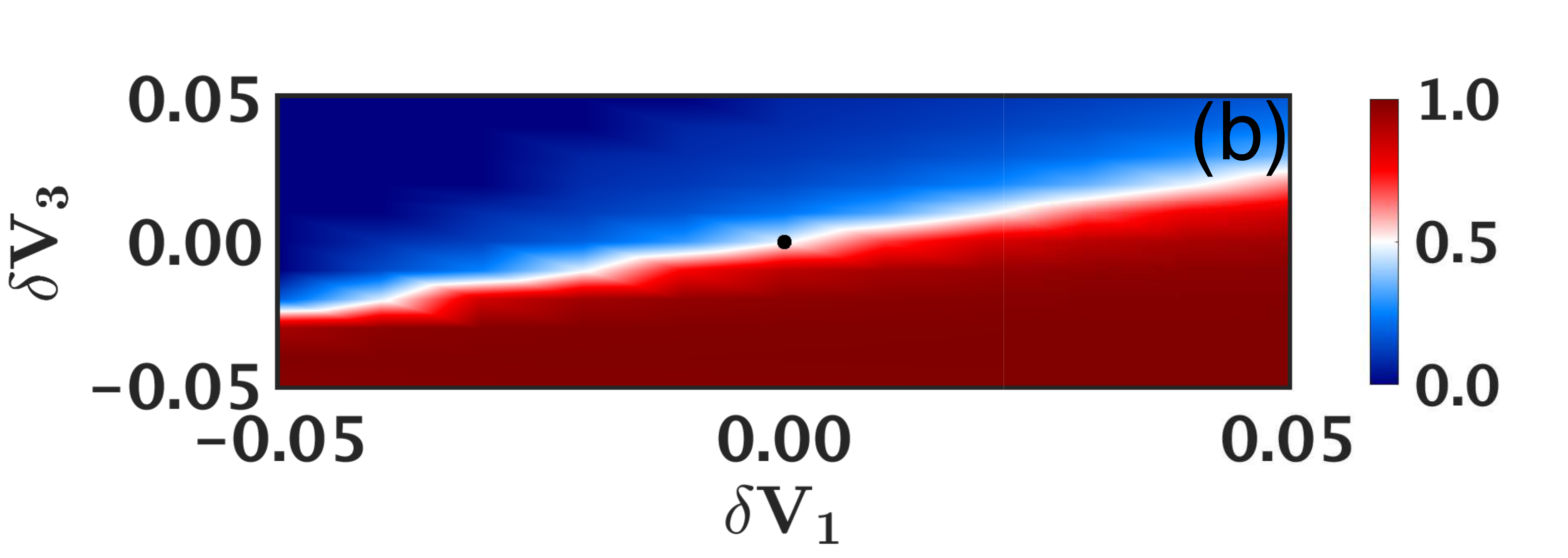} \\
\includegraphics[width=0.47\textwidth,height=0.19\textwidth]{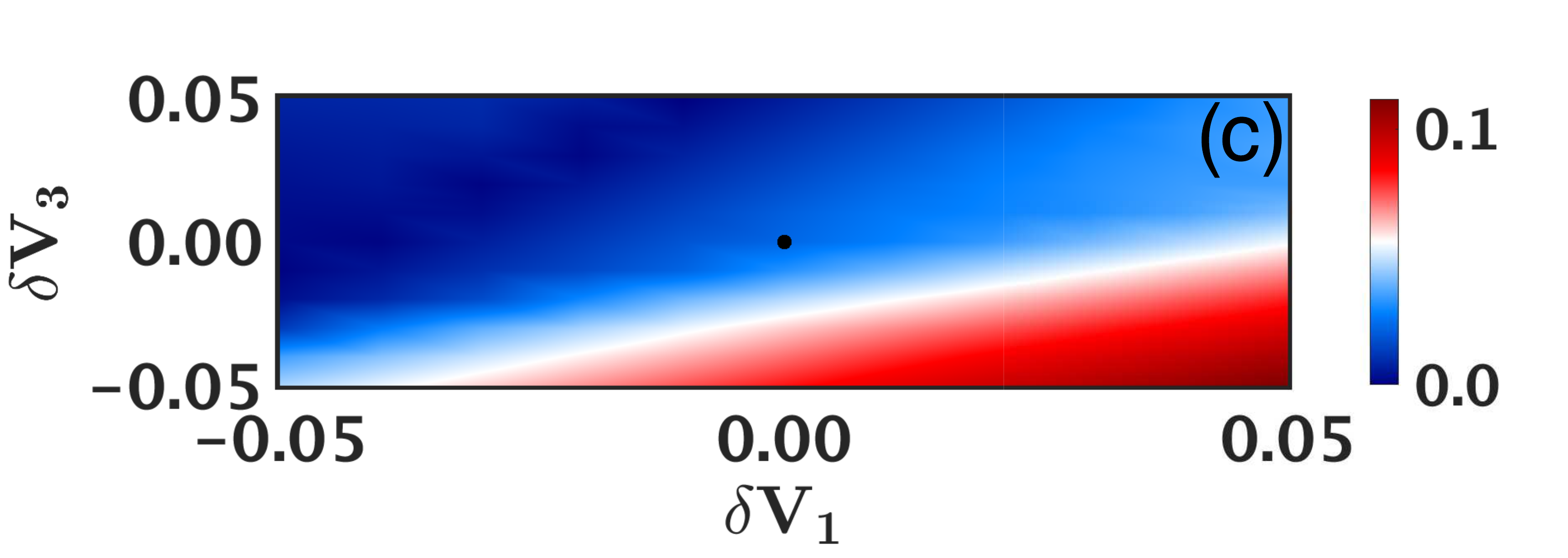} 
\includegraphics[width=0.47\textwidth,height=0.19\textwidth]{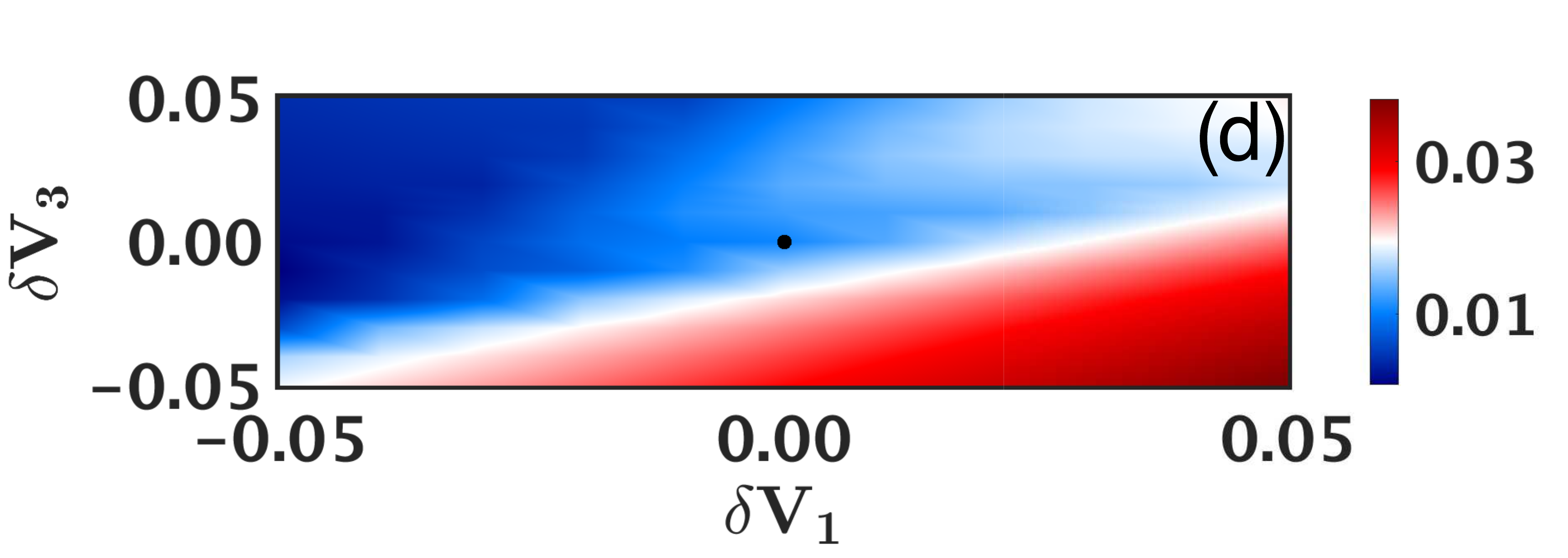}
\caption{(color online) Overlap and density-corrected gap maps obtained in the spherical geometry using exact diagonalization of $N=9$ electrons with the $\bar{3}\bar{3}111$ state at $\nu=3/7$ by perturbing $V_{1}$ and $V_{3}$ around the second Landau level (SLL) spherical pseudopotentials. The system of $N=9$ electrons at a flux of $2Q=24$ aliases with the Laughlin state; therefore, for comparison, in the top right panel we show the overlap map for the $\nu=1/3$ Laughlin state for the same system. The center dot denotes the exact SLL Coulomb point ($V_{1}= 0.4642$ and $V_{3}=0.3635$) for which the values in the four panels are a) $|\langle\Psi^{\bar{3}\bar{3}111}_{3/7}| \Psi^{\rm SLL}_{3/7}\rangle|=0.92$, b) $|\langle\Psi^{\rm Laughlin}_{1/3}| \Psi^{\rm SLL}_{1/3}\rangle|=0.48$, c) neutral gap $\Delta_{n}=0.022$ and d) charge gap $\Delta_{c}=0.010$.}
\label{fig: spectra_bar3bar3111_model_pps_perturb_around_SLL}
\end{center}
\end{figure*}

\begin{figure*}[htpb]
\begin{center}
\includegraphics[width=0.47\textwidth,height=0.19\textwidth]{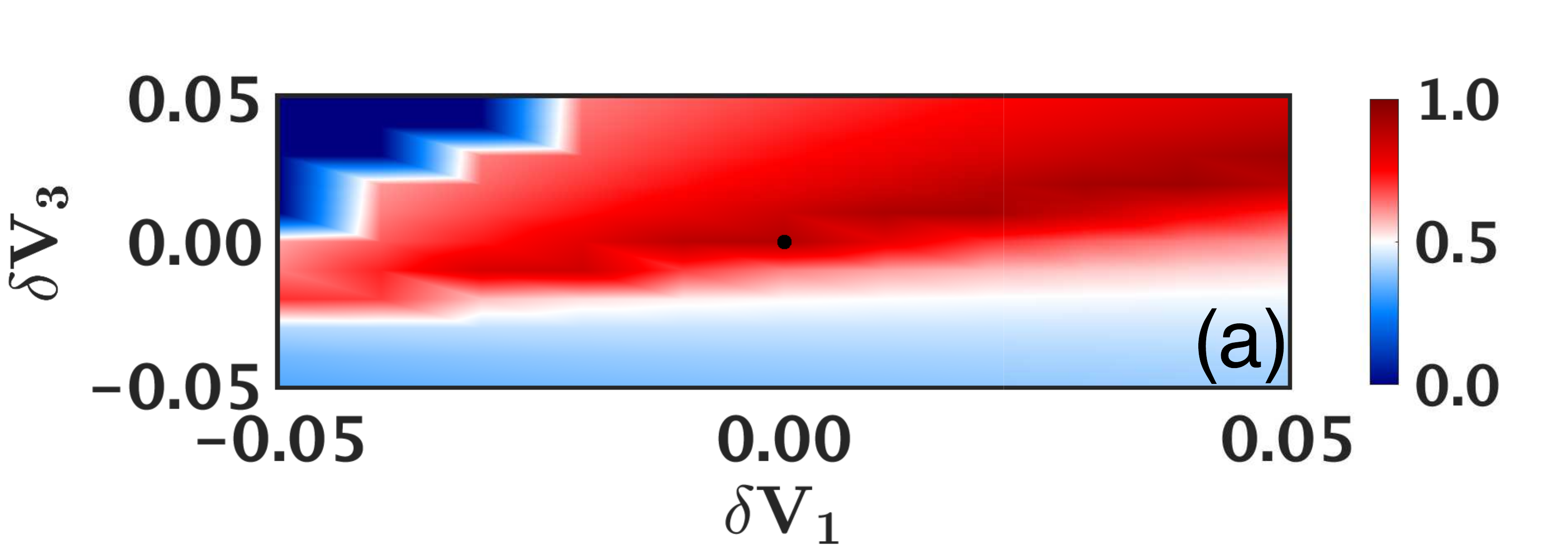}
\includegraphics[width=0.47\textwidth,height=0.19\textwidth]{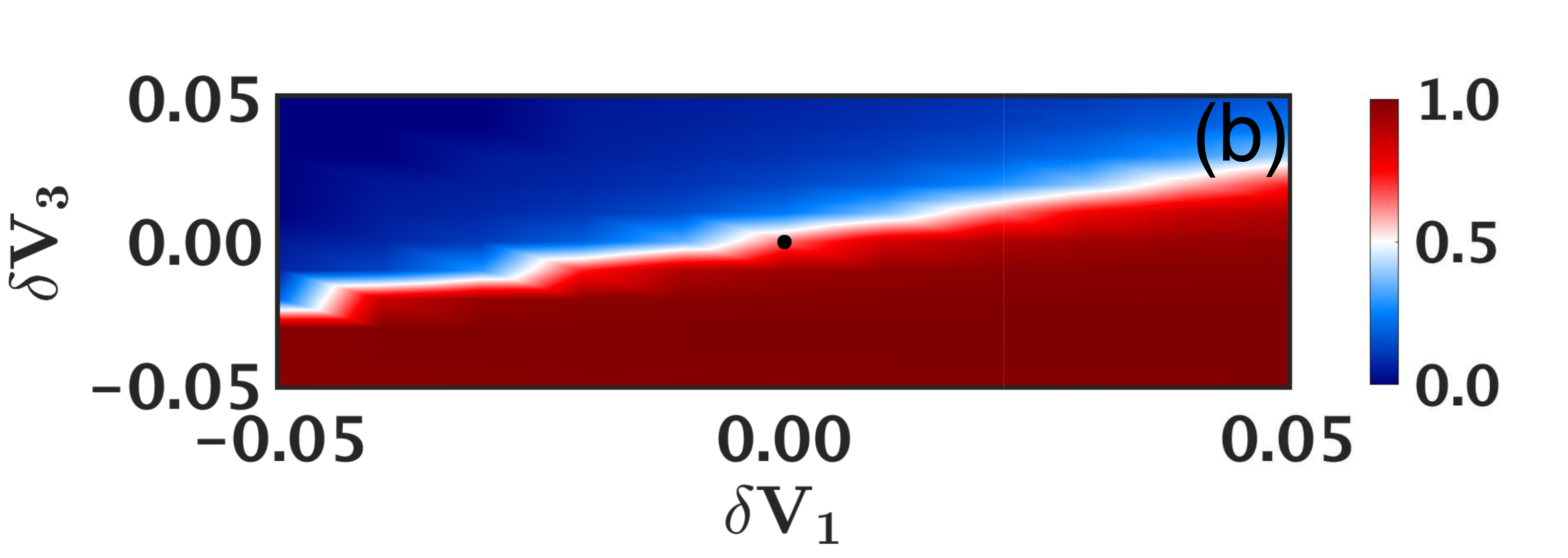} \\
\includegraphics[width=0.47\textwidth,height=0.19\textwidth]{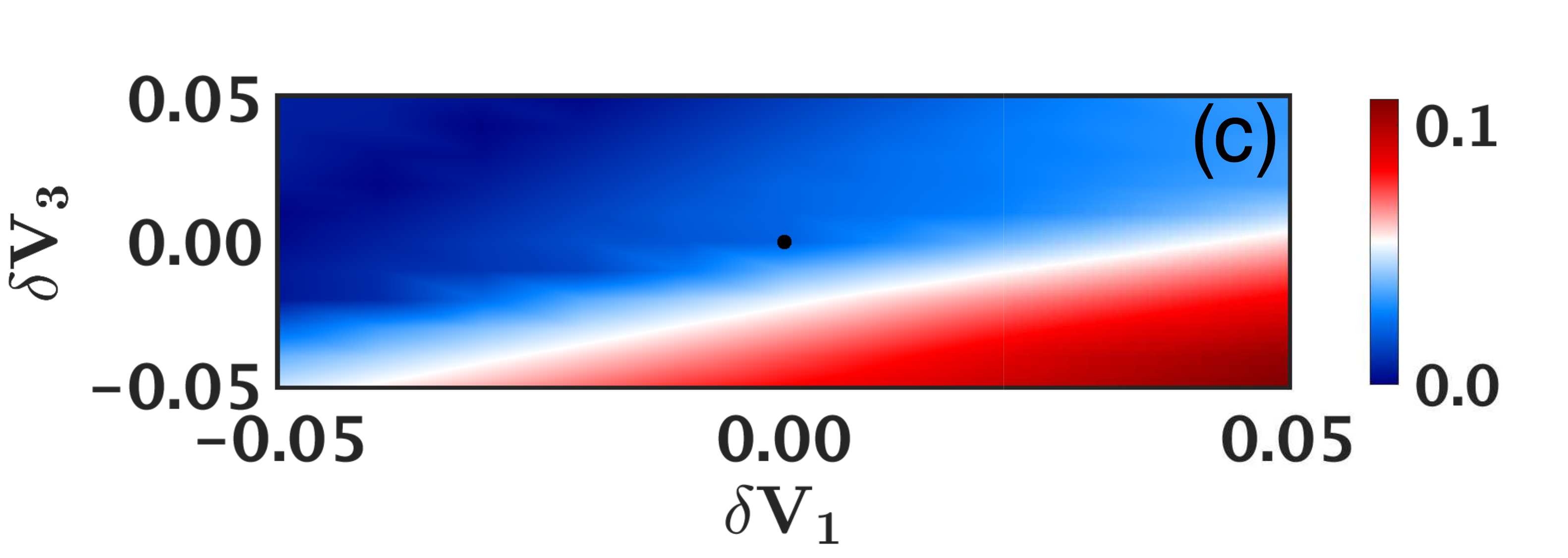} 
\includegraphics[width=0.47\textwidth,height=0.19\textwidth]{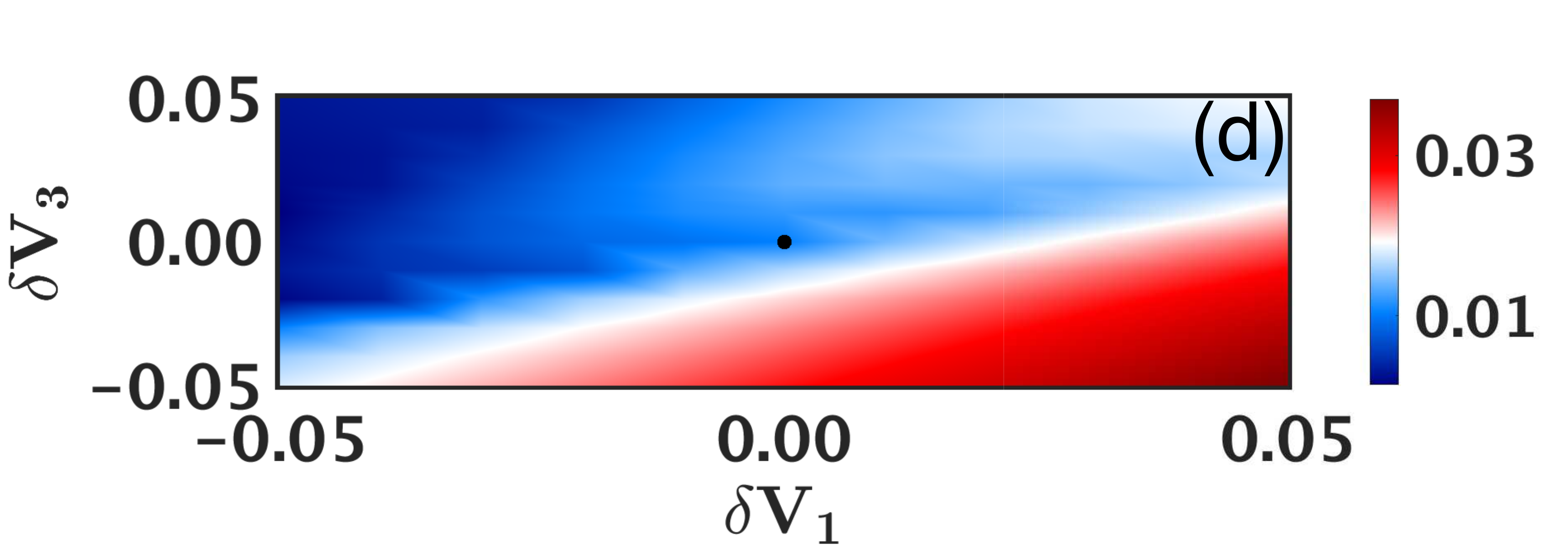}
\caption{(color online) Same as Fig.~\ref{fig: spectra_bar3bar3111_model_pps_perturb_around_SLL} but with results obtained using the second Landau level (SLL) truncated disk pseudopotentials. The center dot denotes the exact SLL Coulomb point ($V_{1}= 0.4154$ and $V_{3}=0.3150$) for which the values in the four panels are $|\langle\Psi^{\bar{3}\bar{3}111}_{3/7}| \Psi^{\rm SLL}_{3/7}\rangle|=0.90$, $|\langle\Psi^{\rm Laughlin}_{1/3}| \Psi^{\rm SLL}_{1/3}\rangle|=0.64$, neutral gap $\Delta_{n}=0.021$ and charge gap $\Delta_{c}=0.010$.}
\label{fig: spectra_bar3bar3111_model_pps_perturb_around_SLL_disk}
\end{center}
\end{figure*}

Next, we test the stability of the $\bar{3}\bar{3}111$ state in the SLL to small perturbations in the interaction. We consider two starting unperturbed interactions: the spherical pseudopotentials and a set of truncated disk pseudopotentials (remember that the interaction is fully defined by the pseudopotentials). We then add small deviations to the $V_1$ and $V_3$ pseudopotentials for each case. We diagonalize these systems to find the charge and neutral gaps, as well as the overlap with the $\bar{3}\bar{3}111$ state. In Figs.~\ref{fig: spectra_bar3bar3111_model_pps_perturb_around_SLL} and \ref{fig: spectra_bar3bar3111_model_pps_perturb_around_SLL_disk}, we present color plots for each quantity in the $\delta V_{1}-\delta V_{3}$ plane for the spherical and truncated disk pseudopotentials respectively. We find that for a wide range of perturbations, the overlap between the $\bar{3}\bar{3}111$ state and the ED ground state remains high, and the state also supports finite neutral and charge gaps. Since for $N=9$ particles the $\bar{3}\bar{3}111$ state occurs at the same flux as the 1/3 Laughlin state, we have also calculated the overlap of the ground states with the Laughlin state. As the upper right panels of Figs.~\ref{fig: spectra_bar3bar3111_model_pps_perturb_around_SLL} and \ref{fig: spectra_bar3bar3111_model_pps_perturb_around_SLL_disk} show, the Laughlin state has a lower overlap than the $\bar{3}\bar{3}111$ state at the SLL Coulomb point, but the overlap grows as the pseudopotentials become more like the LLL, i.e. as $V_3$ decreases relative to $V_1$. 

While the aliasing with the Laughlin state for $N=9$ makes the situation somewhat complicated, the fact that $\bar{3}\bar{3}111$ does better than a very plausible competing state makes it all the more compelling.  As of now, it has not been possible to calculate overlaps for the next system, namely $N=12$, for which we have not been able to obtain the Fock space representation of the $\bar{3}\bar{3}111$ state.

\subsection{Fractional charge of the quasiparticles}

\begin{figure}[htpb]
\begin{center}
\includegraphics[width=0.47\textwidth,height=0.37\textwidth]{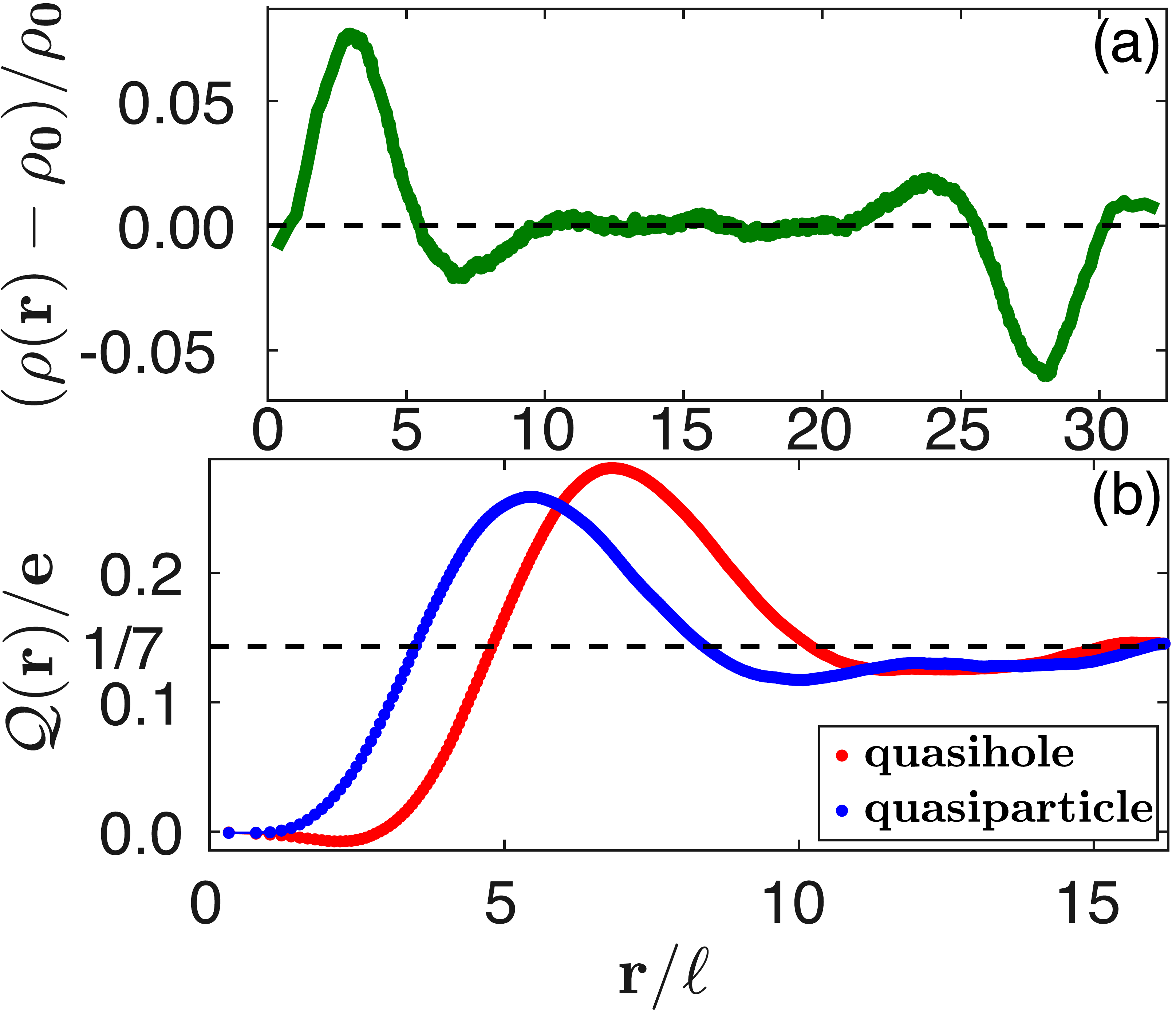}  
\caption{(color online) a) Density profile $\rho(r)$ of a state with a far-separated quasihole and quasiparticle at $\nu=3/7$ modeled by the parton wave function given in Eq.~(\ref{eq: wf_exciton_3_5}) for $N=90$ electrons on the sphere. The quasiparticle is located at the north pole and the quasihole is located at the south pole. The quantity shown is $[\rho(r)-\rho_{0}]/\rho_{0}$, where $\rho_{0}$ is the density of the uniform $\bar{3}\bar{3}111$ state at $\nu=3/7$. b) The quasihole [quasiparticle] cumulative charge $\mathcal{Q}(r)=(-e)\int^{r}_{0} d^{2}\vec{r_{1}}[\rho(\vec{r_{1}})-\rho_{0}]$ [for quasiparticle we show $-\mathcal{Q}(r)$] as a function of the distance $r$ measured along the arc from the south [north] pole to the equator in units of the magnetic length $\ell$. The quasihole [quasiparticle] cumulative charge approaches $-1/7$ [$1/7$], in units of the electron charge $(-e)$, near the equator.}
\label{fig: density_exciton_3bar3bar111}
\end{center}
\end{figure}

The parton theory predicts that the minimal charge quasiparticle of the $\bar{3}\bar{3}111$ state is obtained by creating a quasiparticle in the $\bar{3}$ factor. This excitation carries a charge of $(-e)/7$. It is interesting to note that the non-abelian nature of the $\bar{3}\bar{3}111$ state does not cause a further fractionalization of its quasiparticle charge~(Another example of a non-Abelian state where the charge does not fractionalize further is the $\bar{2}\bar{2}\bar{2}1111$ state at $\nu=2/5$~\cite{Balram19}.). Accessibility to large system sizes allows us to microscopically evaluate the charge of the quasiparticle in the $\bar{3}\bar{3}111$ state. In Fig.~\ref{fig: density_exciton_3bar3bar111} we show the density profile $\rho(r)$ of the $\bar{3}\bar{3}111$ state at $\nu=3/7$ with an exciton where the constituent quasiparticle is located at the north pole and the quasihole at the south pole of the sphere. We consider a system of $N=90$ particles in this calculation. This state is modeled by the wave function given in Eq.~(\ref{eq: wf_exciton_3_5}). Close to the equator the density of the state goes to the density $\rho_{0}$ of the uniform  $\bar{3}\bar{3}111$ state. Note that the density distributions of the quasihole and quasiparticle are not identical since they reside in different $\Lambda$Ls. Furthermore, the quasiparticle and quasihole of $\bar{3}\bar{3}111$ have a wider extent compared to their $311$ counterparts. To demonstrate that the smallest charge quasiparticle (quasihole) of the parton ansatz $\bar{3}\bar{3}111$ has a charge equal to one-seventh of the electron charge, we calculate the integrated cumulative charge $\mathcal{Q}(r) = (-e)\int_{0}^{r} d^{2}\vec{r_{1}}~[\rho(\vec{r_{1}})-\rho_{0}]$ from the north (south) pole to the equator. Doing so for the system of $N=90$ electrons shown in Fig.~\ref{fig: density_exciton_3bar3bar111}, we obtain a charge of magnitude $0.145e$ which is close to the expected value in the thermodynamic limit of $e/7=0.143e$; the small discrepancy arises from the fact that even for a system with $N=90$ particles the quasiparticle and quasihole have a finite overlap. 

%%%%%%%%%%%%%%%%%%%%%%%%%%%%%%%%%%%%%%%%%%%%%%%%%%%%%%%%%%%%%%%%%%%%%%%%%
%%%%%%%%%%%%%%%%%%%%%%%%%%%%%%%%%%%%%%%%%%%%%%%%%%%%%%%%%%%%%%%%%%%%%%%%%

\subsection{Effect of finite well-width}
\begin{figure*}[htpb]
\begin{center}
\includegraphics[width=0.49\textwidth,height = 0.3\textwidth]{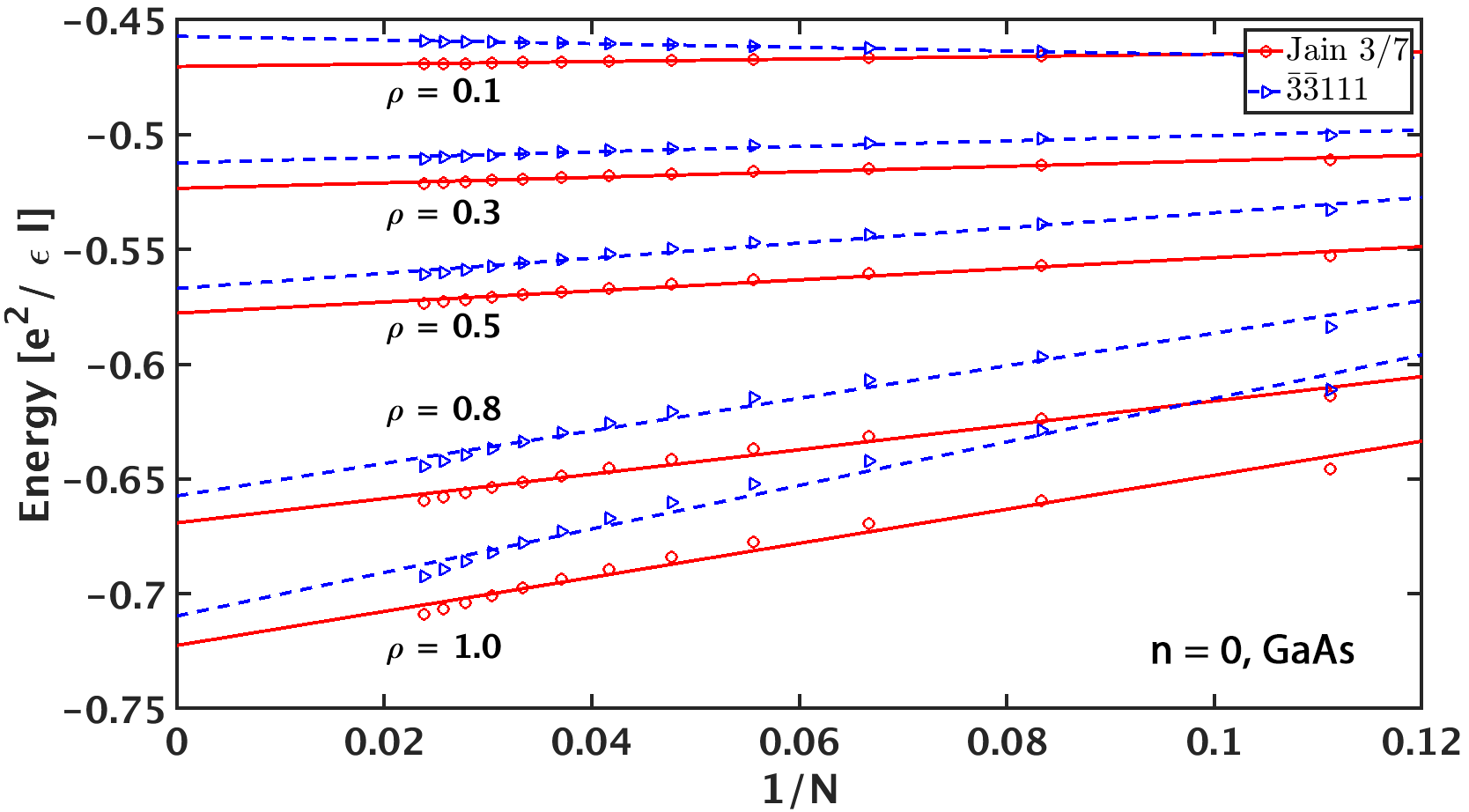}
\includegraphics[width=0.49\textwidth,height = 0.3\textwidth]{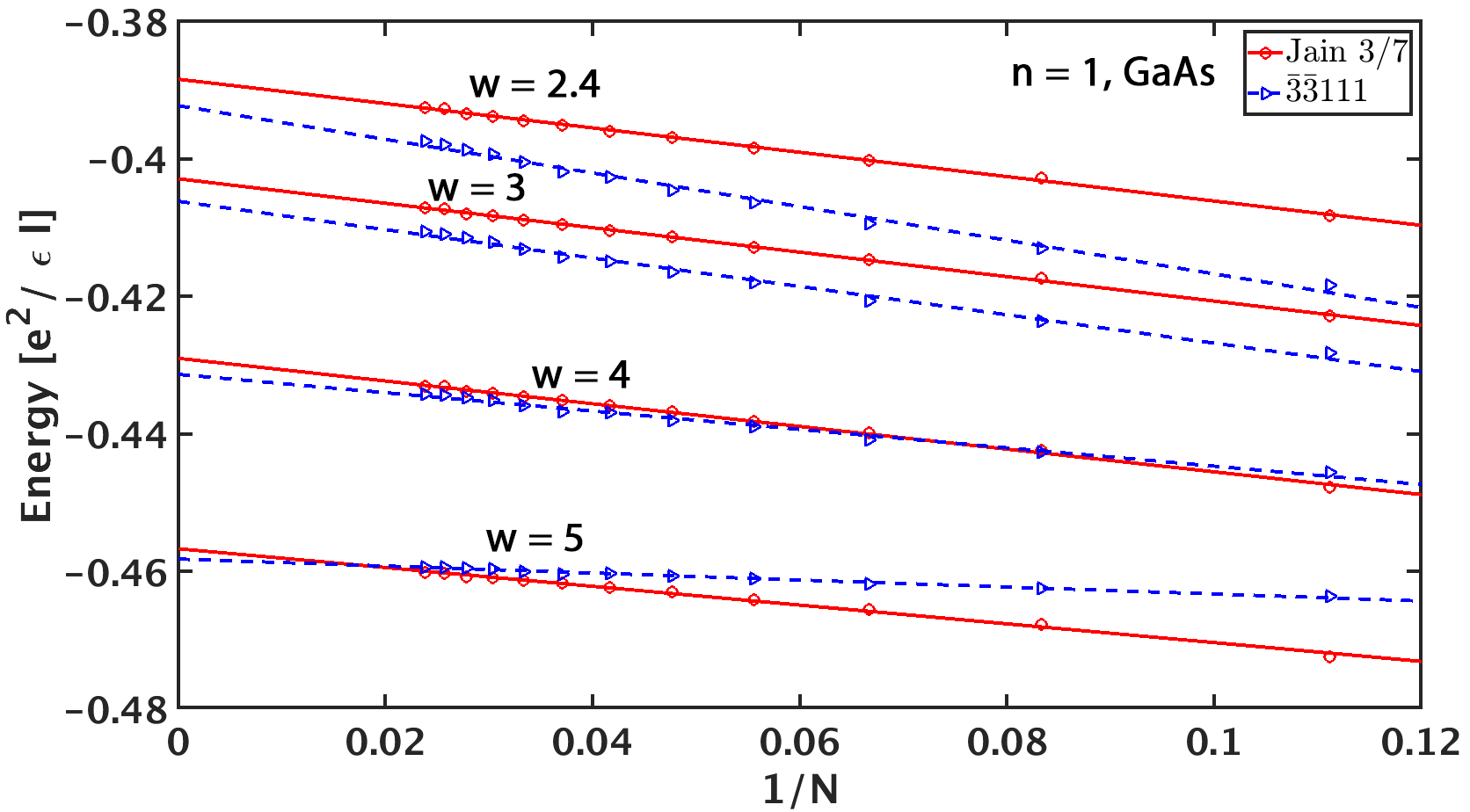}
\caption{(color online) Thermodynamic limits for several finite width systems in the lowest Landau level (LLL) [left panel] and the second LL [right panel]. The LLL results are for a 70 nm well for several densities $\rho$ quoted in units of 10$^{11}$ cm$^{-2}$ on the plot. The second LL finite width interaction is parameterized only by the width of the well $w$ in units of magnetic length.}
\label{fig: finitewidth_TL}
\end{center}
\end{figure*}

We have also considered the effect of the finite well-width $w$ of the quantum well in the LLL and SLL of GaAs. The LLL finite width interaction is obtained via a self-consistent LDA method for a given well width and electron density~\cite{Park99b}. We explore parameters ranging from $w=$18 to 70 nm and electron densities, $\rho$, ranging from 0.1$\times10^{11}$ cm$^{-2}$ to 3$\times10^{11}$ cm$^{-2}$. Representative thermodynamic extrapolations for several different densities at the fixed quantum well width of 70 nm are shown in the left panel of Fig.~\ref{fig: finitewidth_TL}. The Jain $311$ state has lower energy in the LLL for all parameters we have studied.

In the SLL, we use an effective interaction parameterized by the well-width in units of magnetic length for an infinite square well confinement potential, described in detail by T\"oke {\em et al.}~\cite{Toke08}. (Well widths considered correspond to those presented in Table I of \cite{Toke08}). We have calculated energies for well-widths, $w$, up to 5$\ell$, shown in the right column of Fig.~\ref{fig: finitewidth_TL}. We find that the $\bar{3}\bar{3}111$ state has lower energy in the SLL for the entire range of widths considered. A better LDA treatment of the finite width in the SLL is possible and will be needed should experiments observe a phase transition as a function of width; we have not pursued that in this article. 

\section{Discussion and Experimental ramifications}

Our exact diagonalization studies in the spherical geometry for systems with up to $N=15$ particles are consistent with a non-Abelian FQHE at $\nu=2+3/7$. We note that competition between an FQHE liquid and a charge density wave state such as a Wigner or a bubble crystal can be quite subtle~\cite{Zuo20}. The spherical geometry favors a liquid for finite systems, and a crystal may be stabilized in the thermodynamic limit. However, our calculations make a strong case that the $\bar{3}\bar{3}111$ is very competitive here, and {\it if} an FQHE is seen at $\nu=2+3/7$, it likely represents a new non-Abelian state.

One may ask how the $\bar{3}\bar{3}111$ state may be distinguished from the $311$ state. We end the article with a comparison of the various topological properties of the two states. 

The minimally charged quasiparticle carries a charge of magnitude $e/7$ for both the $\bar{3}\bar{3}111$ and the $311$ states. However, the quasiparticles of the $\bar{3}\bar{3}111$ state obey non-Abelian braid statistics~\cite{Wen91}, in contrast to the Abelian quasiparticles of the $311$ state. 

The Hall viscosity of FQH states is expected to be quantized~\cite{Read09}: $\eta_{H} = \hbar \rho_{0} \mathcal{S}/4$, where $\rho_{0}=(3/7)/(2\pi \ell^{2})$ is the electron density and $\mathcal{S}=-3$ is the shift of the $\bar{3}\bar{3}111$ state. In contrast, the Hall viscosity of the $311$ state is given by $\eta_{H} = (5/4)\hbar \rho_{0}$, corresponding to shift ${\cal S}=5$.

Due to the presence of the $\bar{3}$ factors, the $\bar{3}\bar{3}111$ state possesses upstream neutral modes that can be detected in shot noise experiments~\cite{Bid10, Dolev11, Gross12, Inoue14}. Recently, thermal Hall measurements have been carried out at many filling factors in the lowest as well as the second LL of both GaAs~\cite{Banerjee17, Banerjee17b} and monolayer graphene~\cite{Srivastav19}. These thermal Hall experiments can distinguish between the $\bar{3}\bar{3}111$ and $311$ states. In particular, the thermal Hall conductance $\kappa_{xy}$ of the $\bar{3}\bar{3}111$ state is $-(11/5)[\pi^2 k_{\rm B}^2 /(3h)]T$ which is different from what one would expect from the $3/7$ CF state, which has $\kappa_{xy} =3[\pi^2 k_{\rm B}^2 /(3h)]T$ (Note that filled LLs provide an additional contribution to the thermal Hall conductance.).  

The parton state at $\nu = 3/7$ suggests the sequence of $\bar{n}\bar{n}111$ states described by the wave functions
\begin{equation}
\Psi^{\bar{n}\bar{n}111}_{\nu=n/(3n-2)}=\mathcal{P}_{\rm LLL}  [\Phi_{\bar{n}}]^{2}\Phi_1^3 = \frac{[\Psi^{\rm CF}_{n/(2n-1)}]^{2}}{\Phi_{1}}.
\label{eq: parton_barnbarn111} 
\end{equation}
The $\bar{n}\bar{n}111$ state occurs at filling factor $n/(3n-2)$ and has a shift of $\mathcal{S}=3-2n$. The $n=1$ member lies in the same universality class as the standard $\nu=1$ IQHE~\cite{Balram16b}. The $n=2$ member describes a wave function that lies in the same topological phase as the $\nu=1/2$ anti-Pfaffian state~\cite{Balram18}. We considered the $n=3$ member in detail in this work. The $n=4,5,\cdots$ members provide wave functions at $\nu=2/5,5/13,\cdots$ which might be relevant for certain interactions. We leave a detailed exploration of their properties to future work.

We have not performed an energetic comparison with the bubble crystal state. We have not investigated whether the $\bar{3}\bar{3}111$ state at $\nu=3/7$ may occur in a higher LL of monolayer graphene. The state is unlikely to occur in the $n=0$ and $n=1$ LLs of monolayer graphene, whose physics is well described in terms of weakly interacting composite fermions~\cite{Balram15c}, through observation of the standard Jain states at $\nu=n/(2pn\pm 1)$ (see, for example, Refs.~\cite{Feldman13,Amet15,Kim19}). In the $n=2$ LL of monolayer graphene, FQHE has been seen~\cite{Kim19} at $\nu=1/5$ and $\nu=2/9$. It is possible that the $\bar{3}\bar{3}111$ state may occur in $n=2$ or a higher LL of monolayer graphene.

Before closing, we mention other candidate states at $3/7$. 

A non-Abelian candidate state was put forth by Jolicoeur~\cite{Jolicoeur07}. The Jolicoeur wave function is given by:
\begin{equation}
 \Psi^{\rm Jolicoeur}_{3/7}=\mathcal{P}_{\rm LLL} [\Psi^{\rm bosonic-RR}_{3/2}]^{*}\Phi_{1}^{3}, 
 \label{Jolicoeur_wf_3_7}
\end{equation}
where the bosonic version of the three-cluster Read-Rezayi (RR) state~\cite{Read99} is defined as:
\begin{equation}
\Psi^{\rm bosonic-RR}_{3/2}=\mathbb{S}\left[\prod_{l=1,2,3}\prod_{i_{l}<j_{l}}(z_{i_{l}}-z_{j_{l}})^2 \right].
\end{equation}
Here $\mathbb{S}$ denotes the operation of symmetrization of the $N$ particles into three clusters of $N/3$ particles each. The Jolicoeur $3/7$ wave function has a shift of $\mathcal{S}=1$, which is different from our parton state and therefore represents a different topological phase than the parton state. The Jolicoeur wave function is not easily amenable to a numerical calculation and thus we have not considered it in our work. 

Hermanns's hierarchical construction also produces a non-Abelian state at $\nu=4/7$~\cite{Hermanns10}. This state is described by an $su(3)_{2}/u(1)^{2}$ conformal field theory (CFT)~\cite{Hermanns10,Hansson17}. The particle-hole conjugate of this state occurs at $\nu=3/7$ and a shift $\mathcal{S}=-3$ which is the same as the parton state. Just like the Jolicoeur state, the microscopic wave function of the Hermanns's hierarchical state is not readily amenable to a numerical calculation to allow us to compare energies and overlaps. 

Sreejith~\emph{et al}. considered a bipartite CF state at 4/7~\cite{Sreejith11b} which likely lies in the same universality class as the Hermanns hierarchy state. Sreejith~\emph{et al}. showed that the bipartite CF state provides a good representation of the exact SLL Coulomb ground state. These results suggest that the particle-hole conjugate of the $4/7$ Hermanns's hierarchical state is possibly in the same universality class as our parton state. 

Simon, Rezayi and Regnault~\cite{Simon10} constructed a family of wave functions at $\nu=3/7$ based on $S_{3}$ CFTs. These states occur at a shift $\mathcal{S}=5$ which is different from our parton state, and therefore represent a different universality class than our parton state. 

In summary, we have considered the feasibility of the non-Abelian $\bar{3}\bar{3}111$ state for FQHE at $\nu=2+3/7$, i.e.  when the second Landau level is $3/7$ occupied. We have shown that this state has lower energy than the Abelian $311$ state, and also has a high overlap with the exact ground state for small systems. We have also proposed experimental measurements that can reveal the underlying non-Abelian topological order of the $\bar{3}\bar{3}111$ state and distinguish it from the Abelian $311$ state. 

\begin{acknowledgments}
WNF and JKJ are grateful for financial support from the U.S. Department of Energy, Office of Science, Basic Energy Sciences, under Award No. DE-SC0005042. WNF is grateful to the Chateaubriand Fellowship of the Offices for Science and Technology of the Embassy of France in the United States, and to Thierry Jolicoeur for discussions and hospitality at CNRS. We acknowledge useful discussions with Maissam Barkeshli, Ke Huang, Arkadiusz W\'ojs, Ying-Hai Wu, and Jun Zhu. A part of this research was conducted with Advanced CyberInfrastructure computational resources provided by The Institute for CyberScience at The Pennsylvania State University and the Nandadevi supercomputer, which is maintained and supported by the Institute of Mathematical Science’s High-Performance Computing Center. Some of the numerical calculations were performed using the DiagHam package, for which we are grateful to its authors. 
\end{acknowledgments}

%\bibliography{../../Latex-Revtex-etc./biblio_fqhe}
%\bibliographystyle{apsrev}

\end{document}